\documentclass[12pt]{article}
\usepackage{graphicx}
\usepackage{amsmath}
\usepackage{hyperref}
\usepackage{float}
\usepackage{array}
\usepackage[a4paper, total={6in, 8.125in}]{geometry}
\usepackage{multicol}
\usepackage{verbatim}
\usepackage{authblk} 
\emergencystretch=\maxdimen
\providecommand{\keywords}[1]
{
  \small	
  \textbf{Keywords ---} #1
}

\title{Revisiting Stellar Systems J0946 and V723 Monocerotis: A Study of Mass Gap Black Hole Candidates}
\author{
    \small Ajla Trumic$^1$, Aneya Sobalkar$^2$, Efe Tandırlı$^3$, Nishka Yadav$^4$, Isabelle Culinco$^5$,
    \small Shriya Nedumaran$^6$, Kaylee Liu$^7$, Phiet Tran$^8$, Aadhya Pai$^9$, Robert Downing$^{10}$ \\
    \hspace{10pt}
    \small $^1$University of California, Berkeley, CA, 94720, USA \\
    \small $^2$California High School, San Ramon, CA, 94583, USA \\
    \small $^3$Eindhoven Technical University, Eindhoven, North Brabant, 5612, Netherlands \\
    \small $^4$Monta Vista High School, Cupertino, CA, 95014, USA \\
    \small $^5$Campolindo High School, Moraga, CA, 94556, USA \\ 
    \small $^6$Leigh High School, San Jose, CA, 95124, USA \\
    \small $^7$Mission San Jose High School, Fremont, CA, 94539, USA \\
    \small $^8$Santa Clara High School, Santa Clara, CA, 95051, USA \\
    \small $^9$Amador Valley High School, Pleasanton, CA, 94566, USA \\
    \small $^{10}$Aspiring Scholars Directed Research Program, Fremont, CA, 94539, USA
}
\date{}

\begin{document}

\maketitle

\begin{abstract}
\normalsize In 2023, Rowan et al.\textsuperscript{8} reported the discovery of a black hole (BH) companion to J0946, following the misidentification of V723 Mon by Jayasinghe et al.\textsuperscript{5} as containing a “mass-gap” BH. This article reproduced Rowan and Jayasinghe’s results on these systems by estimating stellar parameters via Markov Chain Monte Carlo solvers. We implemented Bayesian statistical modeling through software ExoFit\textsuperscript{3} and PHysics of the Eclipsing Binaries (PHOEBE)\textsuperscript{7}. For J0946, we found a higher inclination ($i\approx72^\circ$) and a companion mass of 2.78 $M_\odot$, lower than what Rowan estimated. V723 Mon’s results aligned with recent estimates by El Badry et al.\textsuperscript{4}, yielding an inclination of $i\approx74^\circ$ and a mass of 2.56 $M_\odot$. We tested this method on stars from Gaia DR2 \textsuperscript{11, 12} and NASA Exoplanet Archive, agreeing with previous findings that these datasets do not exhibit strong indications of stellar-mass black hole systems. \textsuperscript{10}
\end{abstract}

\keywords{mass gap, black holes, PHOEBE, ExoFit, Markov Chain Monte Carlo}

\begin{multicols}{2}
\section{Introduction}
\quad The discovery of neutron stars and black holes (BHs) in the Milky Way galaxy is essential to understanding the end states of stellar evolution. Current mass measurements of neutron stars and black holes are predominantly derived from X-ray detection methodology\textsuperscript{2}. However, only a small percentage of these compact objects are detectable through X-ray methodologies, so these interacting systems are not representative of the total population. To gain a more accurate understanding of how these massive objects form, a more complete census is required. Research into non-interacting binary systems has been crucial in identifying massive companions in the Milky Way galaxy that are indetectable directly by their electromagnetic signatures. The characterization of galactic BH binaries would help bridge the mass gap, defined by the ambiguity between the most massive neutron stars and the least massive black holes in the range of 3-5 $M_\odot$\textsuperscript{5}.

\setlength{\parskip}{0px} 
One possible method of detection is through analyzing stellar radius measurements. Assuming that instrumentation improves over time, it should hold true that stellar radius error measurements will decrease, asymptotically approaching zero. However, in the case of an unseen massive companion in orbit about a common barycenter with the star, the companion’s gravitational attraction will cause tidal distortions that affect the star’s outer envelope. The constant stretching and contraction of the stellar envelope will affect the stellar radius error measurements. These cases, we propose, would present themselves as outliers in radius error measurements and could be filtered into an initial group of candidates. Measurements pertaining to stellar radius were gathered from the 2022 Planetary Systems dataset from the NASA Exoplanet Archive. An additional method of detection is proposed using radial velocity measurements. As the star and the unseen massive companion orbit their barycenter, the star will experience statistically significant deviations in its observed radial velocity and corresponding error measurements due to the influence of the companion. This method relies on the concept that the radial velocity data may reflect that the star is moving in an orbit, suggesting that there may be an unseen companion causing this motion. Gaia astrometry surveys from the European Space Agency provide data pertaining to radial velocity. We, specifically, make use of Gaia Data Release 2 (Gaia DR2)\textsuperscript{11, 12}.

We reproduced the results from D. M. Rowan et al. and Jayasinghe et al. on BH candidate J0946 and imposter V723 Mon, respectively. We test their datasets on the Markov Chain Monte Carlo modeling software, ExoFit, to extract values for period, eccentricity, and radial velocity semi-amplitude. We go on to extract the inclination values through reproducing Bayesian statistical modeling methods from D. M. Rowan et al. through an MCMC built in the PHysics of Eclipsing Binaries (PHOEBE) software. These methods were applied to binary systems identified in our study from the NASA and Gaia datasets. The mass of the unseen companion was then calculated using the binary mass function.

The purpose of this study is to evaluate Rowan and Jayasinghe’s technique on single-lined spectroscopic binaries (SB1s) that could contain black holes that fit in the mass gap range, misidentified BHs, and candidates from our research.

\section{Methodology}
\subsection{Filtering}
\quad To identify any outliers in stellar radius from the NASA Archive, we analyzed parameters: stellar radius (\texttt{st\_rad}), stellar radius error 1 (\texttt{st\_raderr1}), and stellar radius error 2 (\texttt{st\_raderr2}). Null values were replaced with the arithmetic mean of the respective attributes. We identified those stellar bodies whose parameters significantly deviate from the mean using the \texttt{pandas} and \texttt{numpy} libraries. If at least 2 of the 3 parameters described above were found to be statistical outliers, they were sorted into a separate data frame for future analysis. Through leveraging statistical significance and standard deviation thresholds, we identified our first list of potential black holes.

Because Gaia does not include stellar radius measurements, the radial velocity process was applied to Gaia Data Release 2 (Gaia DR2). This dataset contains radial velocity (\texttt{radial\_velocity}) and radial velocity error (\texttt{radial\_velocity\_error}) attributes. Null values were replaced with arithmetic means. We filtered stars that showed significant statistical deviations from their mean in radial velocity through use of the numpy and pandas Python libraries. Anomalous stellar systems were identified and sorted into their own dataset.

To further filter the candidate lists, the datasets produced from the radial velocity and stellar radius methods were merged together into a single data frame using the pandas library. We used right ascension and declination as the shared attribute. The associated error bar was a reference to merge the two datasets. After completion of this process, we were left with a dataset that contained a total of 127 entries and 19 distinct stellar systems.

We then employed the binary mass function to solve for the computed mass of our unseen companion.
\\

Eq. 1: Binary Mass Function
\begin{equation}
f(M) = \frac{(M_2 \sin i)^3}{(M_1 + M_2)^2} = \frac{P_{\text{orb}} K^3 (1-e^2)^{3/2}}{2 \pi G}
\end{equation}
$P_{\text{orb}}$ is the orbital period, $e$ is eccentricity, and $K$ is radial velocity semi-amplitude. $M_1$ and $M_2$ are the masses of the star and the unseen companion, respectively.

To estimate the parameters P\textsubscript{orb}, K, and e, we analyzed various radial velocity time series from a variety of publicly available spectroscopic and photometric datasets.

\section{Data Sources and Processing}
\quad The radial velocity time series was gathered from DACE (Data and Analysis Center for Exoplanets), which provided multiple files for each star from different telescopes and publishing dates. DACE did not contain data for 8 of our stars, so they were removed from our candidates list. 
\end{multicols}
\begin{table}[H]
\centering
\caption{A list of the 11 potential stars, V723 Mon, and J0946 and the radial velocity data sources}
\begin{tabular}{|p{3cm}|p{10cm}|}
\hline
Star Name & Data Source \\ \hline
CoRoT-6 & \href{https://dace.unige.ch/radialVelocities/?pattern=CoRoT-6}{Dace (SOPHIE, pub 2010)} \\ \hline
CoRoT-25 & \href{https://dace.unige.ch/radialVelocities/?pattern=CoRoT-25}{Dace (HARPS pub 2013, HIRES Pub 2013)} \\ \hline
HD 4313 & \href{https://dace.unige.ch/radialVelocities/?pattern=HD%204313}{Dace (HIRES Pub 2010, HIRES Pub 2017)} \\ \hline
HD 37605 & \href{https://dace.unige.ch/radialVelocities/?pattern=HD%2037605}{Dace (HIRES Pub 2017, HRS 2004, HRS 2012, Sandiford Cassegrain Echelle Spectrograph Pub 2004)} \\ \hline
HD 46375 & \href{https://dace.unige.ch/radialVelocities/?pattern=HD%2046375}{Dace (HIRES Post Pub 2021, HIRES Pre Pub 2021, HIRES Pub 2000, HIRES Pub 2006, HIRES Pub 2017)} \\ \hline
HD 128311 & \href{https://dace.unige.ch/radialVelocities/?pattern=HD%20128311}{Dace (HIRES Post Pub 2021, HIRES Pre Pub 2021, HIRES Pub 2003, HIRES Pub 2005, HIRES Pub 2017, HRS Pub 2009)} \\ \hline
HD 171028 & \href{https://dace.unige.ch/radialVelocities/?pattern=HD%20171028}{Dace (HARPS Pub 2007)} \\ \hline
HD 202696 & \href{https://dace.unige.ch/radialVelocities/?pattern=HD%20202696}{Dace (HIRES Pub 2017)} \\ \hline
HD 216520 & \href{https://dace.unige.ch/radialVelocities/?pattern=HD%20216520}{Dace (HIRES Post Pub 2021, HIRES Pre Pub 2021, HIRES Pub 2017)} \\ \hline
TYC 1422-614-1 & \href{https://dace.unige.ch/radialVelocities/?pattern=TYC%201422-614-1}{Dace (HARPN Pub 2015, HARPS-N Pub 2015, HRS Pub 2015)} \\ \hline
WASP-156 & \href{https://dace.unige.ch/radialVelocities/?pattern=wasp-156}{Dace (COR07 DR 3.4, COR14 DR 3.8, SOPHIE Pub 2018)} \\ \hline
V723 Mon\textsuperscript{2} & \href{https://doi.org/10.48550/arXiv.2101.02212}{Jayasinghe et al. 2021} \\ \hline
J0946\textsuperscript{6} & \href{https://gea.esac.esa.int/archive/documentation/FPR/bib.html#bib3}{Gaia Collaboration, Trabucchi, M., et al. 2023c} \\ \hline
\end{tabular}
\end{table}
\begin{multicols}{2}

\setlength{\parskip}{0px}
It was noted that star CoRoT-25 had a discrepancy between radial velocity measurements from different files. Data from HARPS had radial velocities with a mean of -15,427.78 m/s while the data from HIRES had a mean of 1.88 m/s. This issue was fixed by adding an offset of -15.4185 ± 0.007 km/s to the HIRES data to make the measurements consistent.

It was also noted that HD 4313, HD 37605, HD 46375 and HD 128311 had repeated Julian dates. Specifically, each of these stars had multiple measurements with the same Julian date but different radial velocity and error values. This error was disregarded because it did not cause problems in our analysis. Another issue was that the Julian dates for the star TYC 1422-614-1 jumped from an average of 55801.95918 for half of the observations to an average of 105801.95917. The data was split into two datasets, one with an average of 55801.95918 for the Julian dates, and the other with an average of 105801.95917 for the Julian dates. The two datasets were then run in ExoFit and yielded similar results.

Along with radial velocity time series data, a time series for the flux of each star was required. This data was acquired from the ASAS-SN survey\textsuperscript{6, 9}.

\section{Data Analysis}
\quad To provide an initial analysis of our data, we created graphical representations of the radial velocity time series data through PHOEBE. These graphical representations helped us gauge an estimate of our desired parameters. 

To estimate the period, eccentricity, and radial velocity semi-amplitude for each system, we used the software ExoFit. ExoFit is based on the Bayesian statistical technique Monte Carlo Markov Chain (MCMC). We input radial velocity time series data into ExoFit. 

To extract the mass of the unseen companion, it was necessary to estimate the inclination of the binary system. To extract this parameter, we built a model on MCMC using the PHOEBE (Physics of the Eclipsing Binaries) library in Python3. 

To gain an initial estimation for the posteriors, we ran a differential evolution optimizer in PHOEBE for 1,000 iterations on each radial velocity dataset. To reproduce the work of Jayasinghe and Rowan, we set the radius and effective temperature of the unseen companion to 3*10-6 $R_\odot$ and 300 K, respectively, and effects of irradiation and reflection were set to none in the model. We set the uniform and Gaussian priors using the predicted period from ExoFit, effective temperature of the primary component from previous studies used in the NASA Archive, and estimated inclination from the optimizer. The MCMC had 26 walkers and ran for 1,000 iterations, less than that set by Jayasinghe and Rowan. We choose to use less iterations to prove the convergence of walkers to a specific value given fewer iterations. This ensures that the work done by these researchers is reproducible given fewer computational resources. Values for T\textsubscript{eff}, period, and inclination were constrained through Gaussian distributions with loose uniform priors. We created corner plots of our estimated values and plotted our walkers at each iteration to observe the performance of our MCMC. Burnin and thin values were set manually after visually analyzing the walker distribution. 

\section{Results}

\subsection{J0946}
\quad We evaluated our techniques using J0946 which is a giant star predicted to have an unseen, stellar mass BH companion according to recent findings by Rowan et al. We computed J0946’s period, eccentricity, and semi-amplitude of radial velocity parameters using ExoFit.

\begin{figure}[H]
    \centering
    \includegraphics[width=\columnwidth]{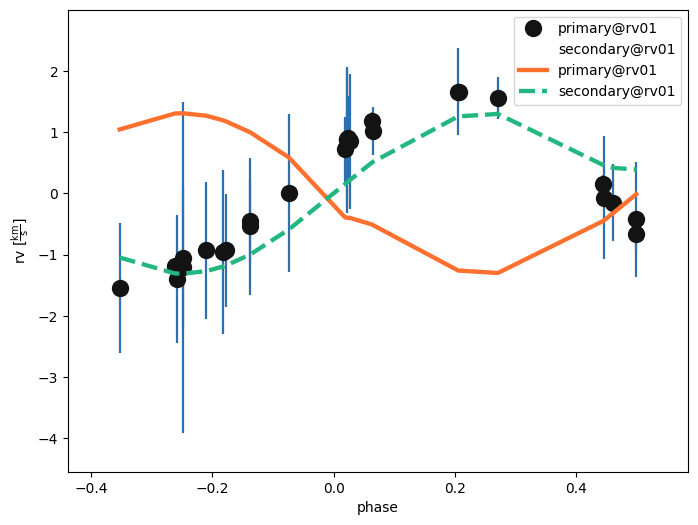}
    \caption{Phase vs. Normalized Radial Velocity for J0946}
    \label{fig:figure1}
\end{figure}

We then used PHOEBE to visualize the phase vs. the normalized radial velocity data in Figure~\ref{fig:figure1}.

\begin{figure}[H]
    \centering
    \includegraphics[width=0.95\columnwidth]{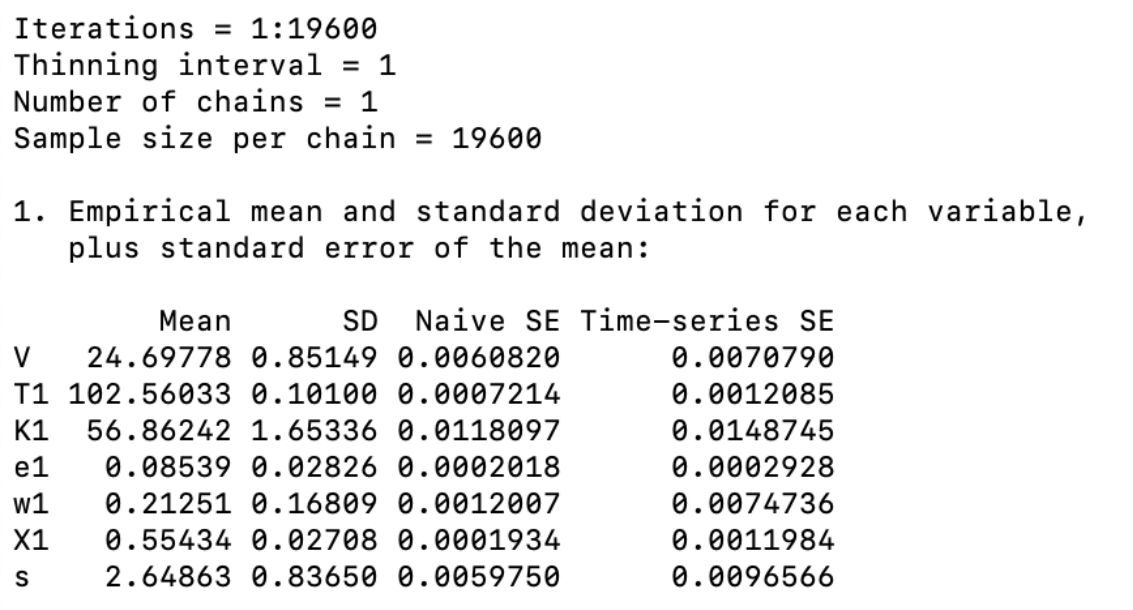}
    \caption{Computed period, eccentricity, and radial velocity semi-amplitude for J0946 from ExoFit. The table displays the mean and standard deviation from ExoFit’s MCMC.}
    \label{fig:figure2}
\end{figure}

\begin{figure}[H]
    \centering
    \includegraphics[width=0.8\columnwidth]{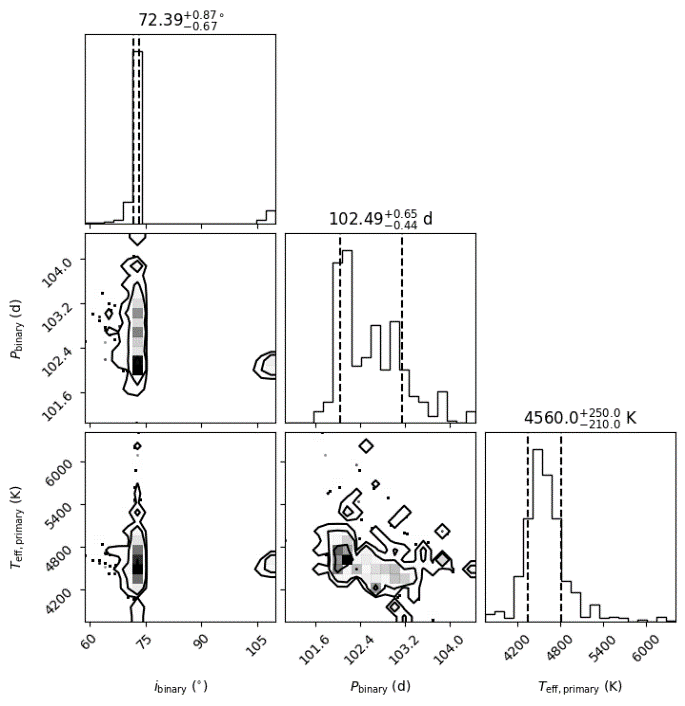}
    \caption{Corner plot showing the estimated inclination, orbital period, and effective temperature of the primary component for J0946. The MCMC ran for 1,000 iterations and twenty-six walkers.}
    \label{fig:figure3}
\end{figure}

The period, eccentricity, and radial velocity semi-amplitude closely matched those found in the paper by Rowan et al. We used the flux and radial velocity data of this star to sample the inclination, period, and effective temperature of the primary component with PHOEBE’s MCMC. The corner plot is shown in Figure~\ref{fig:figure3}.

\begin{figure}[H]
    \centering
    \includegraphics[width=0.5\textwidth]{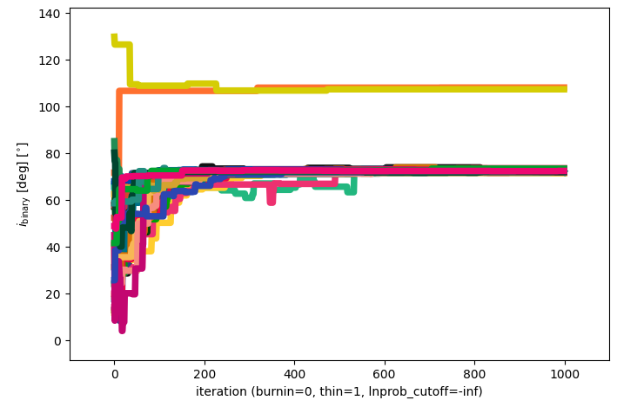}
    \caption{Plot of walkers across 1,000 iterations for J0946. This plot shows the convergence of walkers when estimating the inclination. The MCMC ran for 1,000 iterations and used 26 walkers.}
    \label{fig:figure4}
\end{figure}

The period and temperature of the primary component closely matched that of Rowan et al. However, their posteriors predicted an inclination of $56.05^\circ$ which is less than the $72.39^\circ$ our MCMC posteriors estimated. While it is arguable that our code did not run for as many iterations, we found that our walkers had converged to $72.39^\circ$ after running for only around 600 iterations as seen in Figure~\ref{fig:figure4}.

Each color represents one of the 26 walkers used in Phoebe’s MCMC.

\subsection{V723 Mon}
\quad We evaluated Jayasinghe’s methodology on a misclassified BH candidate to the star V723 Mon to evaluate how robust our methodology is. Initially, V723 Mon was identified as a mass gap black hole candidate but was confirmed by El-Badry et al. to be a low-mass stripped red giant with a subgiant companion. We computed the mean and standard deviation of the estimated period, eccentricity, and radial velocity semi-amplitude using ExoFit (Figure~\ref{fig:figure6}).

\begin{figure}[H]
    \centering
    \includegraphics[width=0.5\textwidth]{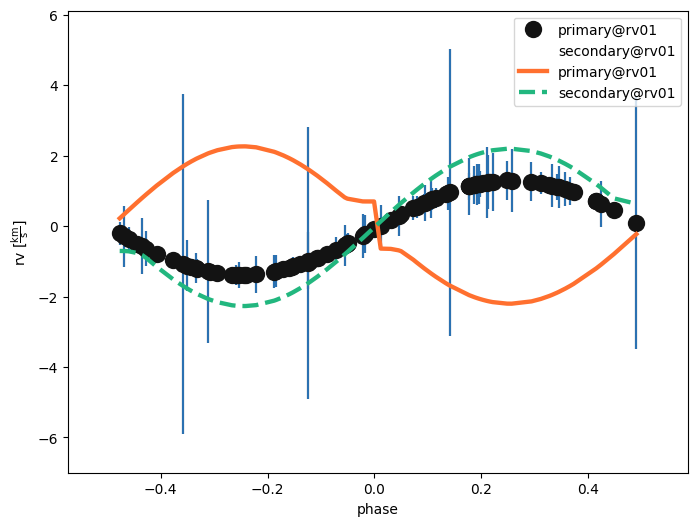}
    \caption{Phase vs. Normalized Radial Velocity for V723 Mon}
    \label{fig:figure5}
\end{figure}

\begin{figure}[H]
    \centering
    \includegraphics[width=0.5\textwidth]{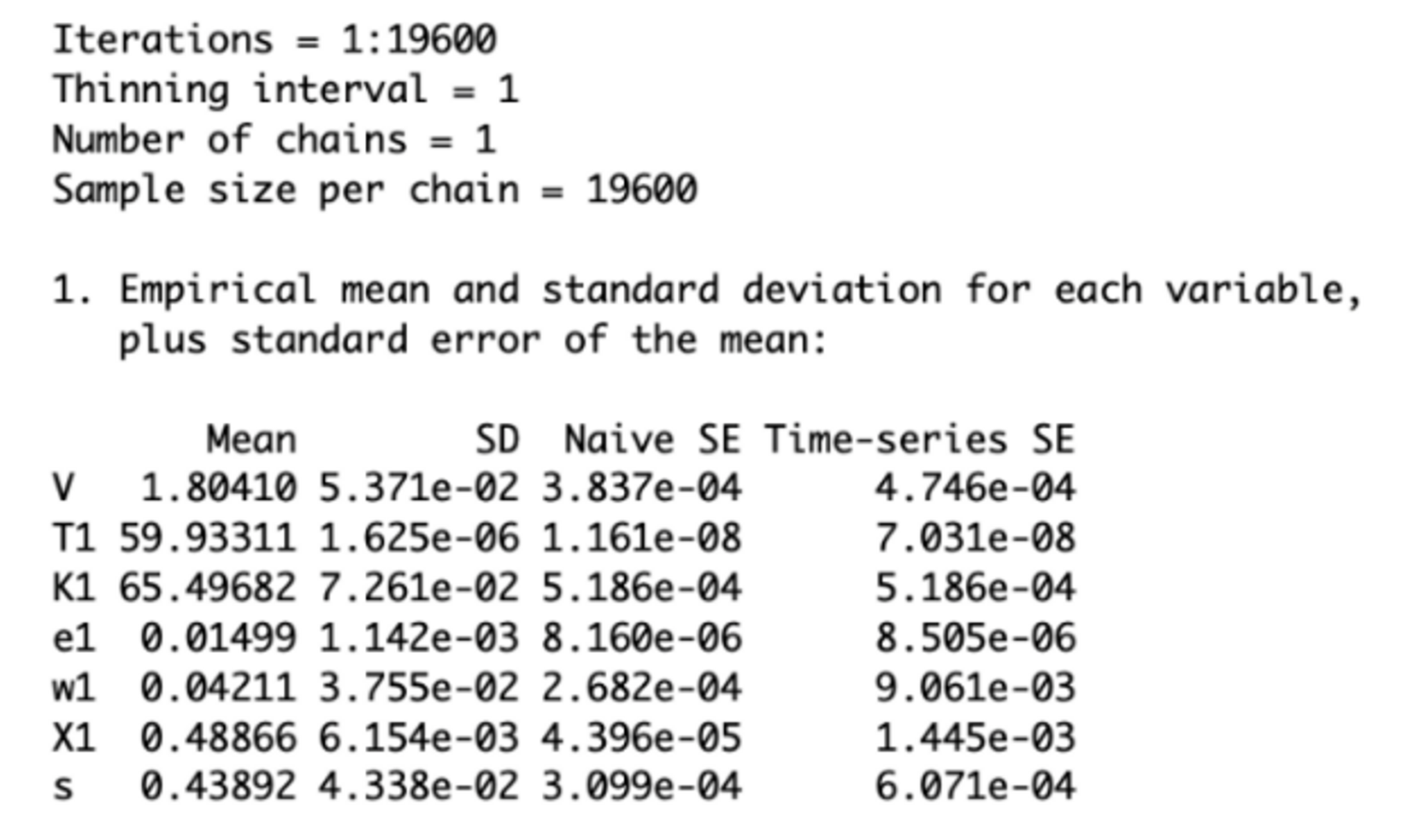}
    \caption{Computed period, eccentricity, and radial velocity semi-amplitude of V723 Mon after running ExoFit. The table displays the mean and standard deviation of estimated parameters.}
    \label{fig:figure6}
\end{figure}

We then used PHOEBE to visualize the phase vs. the normalized radial velocity data in Figure~\ref{fig:figure5}.

We applied the same MCMC solver from PHOEBE to V723 Mon. Using the flux and radial velocity time series data of this star, we sampled the inclination (i\textsubscript{binary}), period (P\textsubscript{binary}), and effective temperature (T\textsubscript{eff, primary}) of the primary component. The corner plots for our results are shown below in Figure~\ref{fig:figure7}.

\begin{figure}[H]
    \centering
    \includegraphics[width=0.5\textwidth]{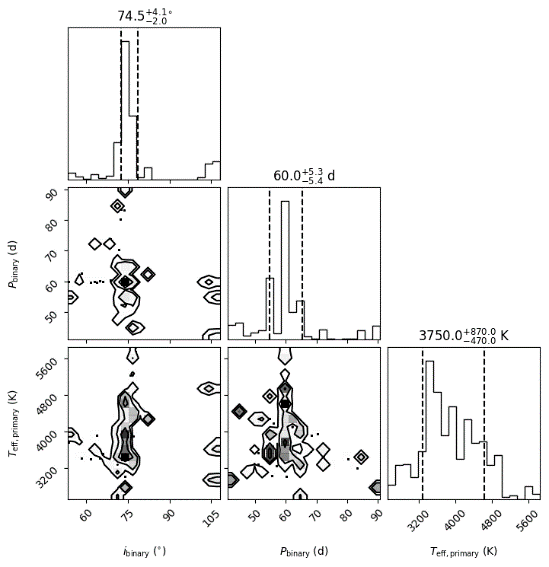}
    \caption{MCMC posteriors for V723 Mon. Corner plot shows the estimated inclination, orbital period, and effective temperature of the primary component. MCMC runs for 1000 iterations and 26 walkers.}
    \label{fig:figure7}
\end{figure}

\begin{figure}[H]
    \centering
    \includegraphics[width=0.5\textwidth]{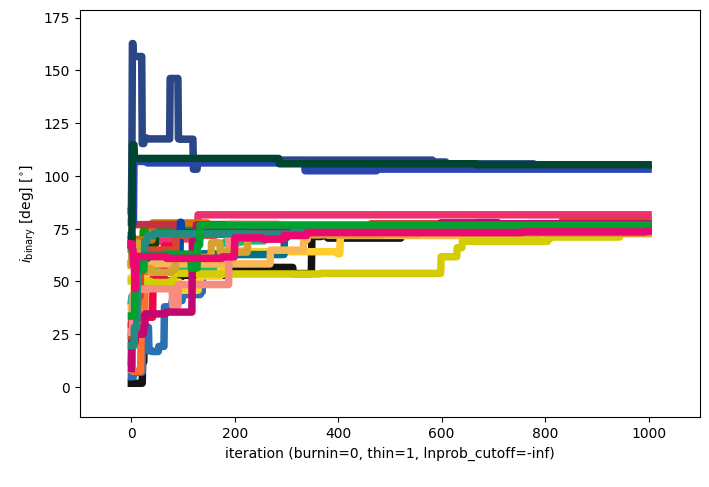}
    \caption{Plot of walkers across 1,000 iterations for V723 Mon. The plot shows the convergence of walkers when estimating the inclination. The MCMC ran for around 800 iterations and used 26 walkers.}
    \label{fig:figure8}
\end{figure}

To evaluate the convergence of the walkers after running the MCMC solver, we plotted the walkers across the iterations to evaluate if running for more iterations would improve the precision of our results. We found that 1,000 iterations sufficed as seen in Figure~\ref{fig:figure8}.

\subsection{Candidate Stars}
\quad After inputting the radial velocity time series for each star in our candidate list in ExoFit, we calculated the inclination angle necessary for an unseen, massive companion that fits in the mass gap range. The values of inclination necessary were not physically plausible nor did they match the values estimated by the differential evolution optimizer. There was minimal convergence by walkers for these stars. There was a strong correlation with the findings from El-Badry \& Rix 2022\textsuperscript{7} in that there are no strong indications of any unseen BHs in this dataset.

\section{Discussion}

\quad For the BH candidate J0946, we computed a mass of $2.78\, M_\odot$ by rearranging the binary mass function to solve for $M_2$. This is slightly less than the previous result\textsuperscript{8}.

We also identified an imposter BH for V723 Mon, with a computed mass of $2.56\, M_\odot$, aligning with the findings from El-Badry et al.\textsuperscript{10} and validating the accuracy of the methods we used. Although our computed mass for V723 Mon was slightly lower than Jayasinghe’s computed mass of $2.8 \pm 0.3\, M_\odot$, our results are still plausible. This is because the inclination angle we found for the system closely aligns with the newest computed range of $62^\circ - 72^\circ$ \textsuperscript{4}. We did not find any mass gap black hole candidates from our candidate list. Our results align with previous examinations of Gaia Data Releases, particularly Gaia DR2 \textsuperscript{1}.

To model radial velocity analytically, we use Equation~\ref{eq:radial_velocity}:

\begin{equation}
    v_i = V - k \left[ \sin(f_i + \omega) + e \sin(\omega) \right]
    \label{eq:radial_velocity}
\end{equation}

$V$ is the systematic radial velocity, $k$ is the semi-amplitude of radial velocity, $\omega$ is the argument of periastron, $f_i$ is the true anomaly, $e$ is the eccentricity, and $v_i$ is the radial velocity.

To estimate our candidates’ parameter values, we used Markov Chain Monte Carlo (MCMC) simulations with the Metropolis-Hastings algorithm in the context of a Bayesian framework\textsuperscript{3}. The MCMC developed in this study addresses significant statistical challenges associated with computing orbital parameters from incomplete or problematic radial velocity data. As a solution, we start with Bayes' theorem, where $\textbf{Y}$ is a vector of $n$ observations with a probability distribution of $P(\theta|\textbf{Y},H)$, where $H$ represents the hypothesis space of the model.

Eq. 3
 \begin{equation} 
     P(\theta|\textbf{Y},H)=\frac{P(\textbf{Y}|\theta, H)P(\theta|H)}{P(\textbf{Y}|H)}
 \end{equation}

The final equation, accounting for the prior distribution, can be rewritten as
\medskip

Eq. 4 

\begin{equation}
    P(\theta|\textbf{Y},H)\propto P(\textbf{Y}|\theta,H)P(\theta,H)
\end{equation}
\setlength{\parskip}{0px}

The likelihood function is crucial in this analysis, modeling the observed radial velocity data while accounting for uncertainties through calculated standard deviation in our datasets. Further, choosing priors significantly impacts the results, emphasizing the need for careful selection. In ExoFit, the choice of priors is preset, and in PHOEBE, priors are chosen according to estimates from the differential evolution optimizer and prior studies. Markov Chain Monte Carlo (MCMC) methods are used to simulate posterior distributions, efficiently handling complex probability models. MCMC uses four key objects within the object-oriented analysis: state, data, bond, and update. State defines the Markov Chain within the program. The parameters within state are then linked by bond, which includes prior densities and likelihood, representing the posterior density without the normalization constant from Bayes’ Theorem (Eq. 4). The Update function then determines which parameters should be adjusted during each iteration, proposing new values and calculating the corresponding bond strength for the proposed state. This new state is either accepted or rejected based on the Metropolis-Hastings method which both ExoFit and PHOEBE used.

According to estimates by Rowan et al. we expected an inclination angle of 50-60\textdegree, but our actual estimation for the inclination angle, i=71\textdegree, exceeded this range. This finding is significant as it suggests that the mass of the unseen, massive companion to J0946 is smaller than expected and is slightly below what would be in the “mass gap” range. This puts previous results into question, which we had aimed to evaluate in this paper. Given the requirements to be an eclipsing binary, it is more likely that the inclination angle would be larger and closer to 90\textdegree. However, since PHOEBE is specifically designed to model and analyze eclipsing binary systems, this method may have caused our computed inclination angle to be higher than that of D. M. Rowan et al.

Though the companion to J0946 has a lower mass than expected, it can still be best categorized as housing an unseen massive companion. Further work will help to more rigorously determine the nature of J0946’s companion to see if it is an imposter system. The V723 Mon binary system was also identified to have a similar mass estimation compared to previous studies.

The results of the filtration process on the NASA Archive and Gaia DR2 data found no values of inclination that would mathematically yield a companion mass in the proposed mass gap range. This aligns with previous findings\textsuperscript{10}. This helps support the validity of Rowan and Jayasinghe’s techniques. 

Potential limitations posed are limited computational resources that affect the number of possible iterations to run. Additionally, a lack of data for certain stars hindered data analysis from some stellar systems.

\section{Conclusion}

\quad Our research addresses the mass gap range for BHs which is crucial in expanding our understanding of galactic BHs and gaining a deeper comprehension of stellar evolution. Our work offers the opportunity to detect BHs through methodology apart from established X-ray or microlensing techniques. Making use of publicly available spectroscopy and photometry data, we reviewed and adjusted BH detection methods to allow for improvements in ongoing efforts to characterize the BHs in the Milky Way Galaxy. 

We reviewed the misidentified black hole companion to the star, V723 Mon, allowing for analysis of “imposter” systems. Our research ensures replicability of the process by Rowan et al. through the review and reproduction of the results for J0946. 

In the future, we plan on expanding our research methodology to new datasets, such as those from the Gaia Early Focused Product Release as well as radial velocity and flux time series data for all stars. Future implementation of MCMC could include expanding PHOEBE’s parameters to better model the system. PHOEBE is set to update its program to include more analyses that fit SB1’s with potential black hole systems. However, the date for this release is currently unknown.

The reproduction of our results with newer versions of PHOEBE may provide a more accurate analysis of these systems. Additionally, increased computational power would allow for more iterations and walkers, which could be adjusted to test the posterior settings and better allow the walkers to explore the parameter space. 

This study also did not use spectral energy distributions (SEDs). To better characterize the unseen massive companion detected, future research may assess this limitation through analysis of spectra data.

\section*{Acknowledgements}

\quad This work has made use of data from the European Space Agency (ESA) mission Gaia (\url{https://www.cosmos.esa.int/gaia}), processed by the Gaia Data Processing and Analysis Consortium (DPAC, \url{https://www.cosmos.esa.int/web/gaia/dpac/consortium}). Funding for the DPAC has been provided by national institutions, in particular the institutions participating in the Gaia Multilateral Agreement.

This publication makes use of The Data \& Analysis Center for Exoplanets (DACE), which is a facility based at the University of Geneva (CH) dedicated to extrasolar planets data visualization, exchange, and analysis. DACE is a platform of the Swiss National Centre of Competence in Research (NCCR) PlanetS, federating the Swiss expertise in Exoplanet research. The DACE platform is available at \url{https://dace.unige.ch}.

This research has made use of the NASA Exoplanet Archive, which is operated by the California Institute of Technology, under contract with the National Aeronautics and Space Administration under the Exoplanet Exploration Program.
\section*{ORCID iDs}
\href{https://orcid.org/0009-0007-9417-9214}{\\Ajla Trumic: 0009-0007-9417-9214}
\href{https://orcid.org/0009-0008-6722-5454}{\\Aneya Sobalkar: 0009-0008-6722-5454}
\href{https://orcid.org/0009-0004-7997-6539}{\\Efe Tandırlı: 0009-0004-7997-6539}
\href{https://orcid.org/0009-0002-8412-7633}{\\Nishka Yadav: 0009-0002-8412-7633}
\href{https://orcid.org/0009-0002-7065-6411}{\\Isabelle Culinco: 0009-0002-7065-6411}
\href{https://orcid.org/0009-0007-2092-7558}{\\Shriya Nedumaran: 0009-0007-2092-7558}
\href{https://orcid.org/0009-0007-2251-5311}{\\Kaylee Liu: 0009-0007-2251-5311}
\href{https://orcid.org/0009-0006-6710-0227}{\\Phiet Tran: 0009-0006-6710-0227}
\href{https://orcid.org/0009-0008-6349-5713}{\\Aadhya Pai: 0009-0008-6349-5713}
\href{https://orcid.org/0000-0003-3967-817X}{\\Robert Downing: 0000-0003-3967-817X}

\end{multicols}


\begin{thebibliography}{12}

\bibitem{Arenou2018} F. Arenou, et al., ``GAIA Data Release 2,'' \textit{Astronomy and Astrophysics}, vol. 616, p. A17, 2018.

\bibitem{Avakyan2023} A. Avakyan, et al., ``XRBcats: Galactic Low Mass X-ray Binary Catalogue,'' 27 July 2023. [Online]. Available: \url{https://arxiv.org/abs/2303.16168}.

\bibitem{Balan2009} S. T. Balan, O. Lahav, ``EXOFIT: Bayesian Estimation of Orbital Parameters of Extrasolar Planets,'' 21 July 2009. [Online]. Available: \url{https://arxiv.org/pdf/0907.3613}.

\bibitem{ElBadry2022} K. El-Badry, et al., ``Unicorns and Giraffes in the binary zoo: stripped giants with subgiant companions,'' \textit{Monthly Notices of the Royal Astronomical Society}, vol. 512, no. 4, pp. 5620–5641, 2022. [Online]. Available: {http://dx.doi.org/10.1093/mnras/stac815}

\bibitem {Jayasinghe2021} T. Jayasinghe, et al., ``A Unicorn in Monoceros: the $3 M_\odot$ dark companion to the bright, nearby red giant V723 Mon is a non-interacting, mass-gap black hole candidate,'' \textit{Monthly Notices of the Royal Astronomical Society}, vol. 504, no. 2, pp. 2577–2602, 2021. [Online]. Available: \url{http://dx.doi.org/10.1093/mnras/stab907}

\bibitem{Kochanek2017} K. Kochanek, et al., ``The All-Sky Automated Survey for Supernovae (ASAS-SN) Light Curve Server v1.0,'' \textit{Publications of the Astronomical Society of the Pacific}, vol. 129, no. 980, 2017. [Online]. Available: \url{http://dx.doi.org/10.1088/1538-3873/aa80d9}

\bibitem{Prsa2011} A. Prsa, et al., ``PHOEBE: PHysics Of Eclipsing BinariEs,'' \textit{Astrophysics Source Code Library}, 2011. [Online]. Available: \url{https://ui.adsabs.harvard.edu/abs/2011ascl.soft06002P/abstract}

\bibitem{Rowan2024} D. Rowan et al., ``High mass function ellipsoidal variables in the Gaia Focused Product Release: searching for black hole candidates in the binary zoo,'' 17 January 2024. [Online]. Available: \url{https://arxiv.org/abs/2401.09531}

\bibitem{Shappee2014} B. J. Shappee, ``The Man behind the Curtain: X-Rays Drive the UV through NIR Variability in the 2013 Active Galactic Nucleus Outburst in NGC 2617,'' \textit{The Astrophysical Journal}, vol. 788, no. 1, 2014. [Online]. Available: \url{http://dx.doi.org/10.1088/0004-637X/788/1/48}

\bibitem{ElBadry2022b} K. El-Badry, H. W. Rix, ``What are the spectroscopic binaries with high-mass functions near the Gaia DR3 main sequence?,'' \textit{Monthly Notices of the Royal Astronomical Society}, vol. 515, no. 1, pp. 1266–1275, 2022. [Online]. Available: \url{http://dx.doi.org/10.1093/mnras/stac1797}

\bibitem{Trabucchi2023} Trabucchi, et al., ``Gaia Focused Product Release: Radial velocity time series of long-period variables,'' \textit{Astronomy and Astrophysics}, vol. 680, p. A36, 2023.

\bibitem{Riello2018} M. Riello, et al., ``GAIA Data Release 2,'' \textit{Astronomy and Astrophysics}, vol. 616, p. A3, 2018.

\end{thebibliography}
\end{document}